\documentclass[12pt,english]{article}

\usepackage[hidelinks]{hyperref}

\usepackage{authblk}

\usepackage{xcolor}

\usepackage{graphicx}
\usepackage{babel}
\usepackage{amsmath}
\usepackage{enumerate} 
\usepackage{float}
\usepackage[utf8]{inputenc}\usepackage[T1]{fontenc}

\usepackage{geometry}
\makeatletter
\newcommand\iraggedright{
 \let\\\@centercr\@rightskip\@flushglue \rightskip\@rightskip
\leftskip\z@skip}
\makeatother
\iraggedright

\usepackage[T1]{fontenc}

\usepackage{txfonts}

\usepackage{setspace}

\usepackage{parskip}
\usepackage[]{footmisc}

\usepackage[font=normalsize,labelfont=bf,justification=raggedright]{caption}

\usepackage{titlesec}
\titleformat*{\section}{\bf}
\titleformat*{\subsection}{\it}
\titleformat*{\subsubsection}{\it}

\begin{document}

\title{On emergence from the perspective of symmetry-breaking in physical science}

\author{Joon-Young Moon}
\affil{Department of Psychological and Brain Sciences, Johns Hopkins University, Baltimore, MD, USA}

\author{Eric LaRock}
\affil{Department of Philosophy, Oakland University, Rochester, MI, USA}

\date{}

\maketitle

\thispagestyle{empty}

\begin{abstract}
In this paper, we summarize the development of the concept of emergence in physical science and propose key concepts of emergence in the form of conjectures related to symmetry-breaking. Our conjectures are threefold: I. A system having a broken-symmetry in membership relation with respect to micro and macro scales can have emergent properties. II. Spontaneous symmetry-breaking is an example of an emergent property. III. The phenomenon of hysteresis accompanies spontaneous symmetry-breaking. We argue that these conjectures and their relationship can illuminate the concept of emergence from the perspective of symmetry breaking. 
 \\[12pt]
\end{abstract}

\pagebreak

\section{Introduction and concept of emergence}

A property of a system is said to be \textit{emergent} if it is in some sense more than the sum of the properties of the system's parts.  \textit{Emergence} involves a process that exemplifies a higher-level, global property in relation to lower-level, local properties.  Framed in terms of parts and wholes, one might say, along with Bar-Yam (1997), that "what parts of a system do together" would not be done alone were it not for higher-level, global properties.  Thus an emergent (or higher-level, global) property, in our sense, makes a causal (and thus explanatory) difference to the way lower-level, local properties interact and function together.

\section{Statement of the purpose}

The purpose of this paper is to summarize the development of the concept of \textit{emergence} in the field of physical science, with a few original propositions from the authors. Where the new propositions appear, we specify. We will formalize propositions (some old and some new) in a form of conjectures. These propositions will look at the concept of emergence from the perspective of symmetry-asymmetry relationships. Towards the end, we will summarize our endeavor and state three conjectures as the crucial propositions on the concept of emergence. These critical propositions of emergence can all be written as the symmetry-asymmetry relationships of various physical properties of the system such as membership relationships, and time reversal relationships. Our view is related to experimental examples in physics and constitutes a further theoretical study with respect to the  \textit{New Emergentist Thesis} advocated by Mainwood (2006).  We begin with a description of the concept of emergence.

\section{Brief history: Anderson's paper}

Modern usage of the concept of emergence began among philosophers and can be traced back to John Stuart Mill (1843) and George Henry Lewes (1875).  The concept gained traction among philosophers, the so-called \textit{British Emergentism}, and eventually culminated in Charlie Dunbar Broad's (1925) seminal work, \textit{Mind and Its Place in Nature}.  However, in the field of science, with the success of the \textit{reductionistic approach} throughout the 20th century, the emergentist view slowly retracted from the fields of physics, chemistry, and biology (McLaughlin 1992). As Mainwood observes:

 \begin{quotation}
\noindent "The sciences of the very small (for example, high energy particle physics) would be the one engaged in a search for fundamental laws, and other sciences could be in a sense derivative, in that their subject matters are ultimately the behavior of physical systems and this could be derived in theory from the laws and principles that governed the very small." (Mainwood 2006, 3-4)
 \end{quotation}

However, a significant change took place in 1972 that put emergentism back on the theoretical table, when Nobel Prize recipient condensed matter physicist Philip Anderson published a short paper, \textit{More is Different}.  Anderson advocated the foundational concepts of emergentism in a pithy way as follows:

\begin{quotation}
\noindent "At each level of complexity entirely new properties appear\dots At each stage entirely new laws, concepts, and generalizations are necessary\dots Psychology is not applied biology, nor is biology applied chemistry. We can now see that the whole becomes not merely more, but very different from the sum of its parts." (Anderson 1972, 393)
\end{quotation}

Anderson's paper rekindled the discussion of emergence once again among scientists, now summarized as the \textit{New Emergentist Thesis}; and it is on these discussions that we focus our attention here.

\section{Thesis A: broken symmetry in membership relation and emergence}

Anderson (1972) motivates the novelty of emergent properties within a hierarchy of sciences, according to the idea that the elementary entities of science X obey the laws of science Y.

\begin{center}
\begin{tabular}{*{2}{c}}
X & Y\\
\hline
Solid state/many-body physics & Elementary particle physics\\
Chemistry & Many-body physics\\
Molecular biology & Chemistry\\
Cell biology	& Molecular biology\\
. & . \\
. & . \\
. & . \\
Psychology & Physiology\\
Social sciences	 & Psychology\\
\end{tabular}
\end{center}

Then, Anderson goes on to state that "this hierarchy does not imply that science X is `just applied' science X".

\begin{quotation}
\noindent "At each stage entirely new laws, concepts, and generalizations are necessary, requiring inspiration and creativity to just as great a degree as in the previous one." (393)
\end{quotation}

If a macro-level, systemic property, X, is truly novel, then X and its governing laws could be regarded as exemplifying a reality status that is just as great as its micro-level properties and laws. Mainwood makes the following useful remark in this context: "some systemic properties are importantly novel: so different to the microphysical that they and the laws that govern them can be recognized as having a metaphysical status in no way inferior to the microphysical." (20) 

For example, there exists a fundamental \textit{broken-symmetry}, in the sense that parts in level Y are subsets of the whole system (at level X), but the whole system is not a subset of the parts in level Y. We name such a property \textit{hierarchical realism} after Christen et al. (2002).  In what follows, we conjoin hierarchical realism and novel emergent properties.  We formally define those notions as follows: 

\begin{enumerate}[I.]
\item Hierarchical realism: The system under consideration has at least two hierarchical levels, a higher level X where the system itself exists, and a lower level Y where the parts making up that system exists. There exists a broken-symmetry between the two levels, in the sense that the parts at Y are subsets of the whole system, but the whole system is not the subset of the parts. We can call this property the \textit{broken-symmetry in membership relation}.

\item Novelty of emergent properties: Some properties/laws at the system level are novel, and different from those at the micro-level, such that they have the same reality status as the micro-level properties/laws.
\end{enumerate}

Following Anderson, we maintain that novel properties exist at the macro level.  This claim is at the heart the \textit{New Emergentist Thesis}. Whether this claim is achievable will depend on its testability. We may restate the claim as a form of conjecture:

\begin{quotation}
\noindent \textbf{Conjecture A (New Emergentist Thesis)}. A system realizing hierarchical realism (having a broken-symmetry in membership relation) can have novel emergent properties at the macro level.
\end{quotation}

We note that if we denote the concept of broken-symmetry in membership-relation in set theoretical terms, we meet with the forbidden realm of modern set theory (Jech 2006). In naive set theoretic notations, we can write the broken-symmetry in membership relation as the following:
\begin{equation*}
\begin{split}
M= \{&m_1,m_2,m_3,\ldots,m_n\}, \\
here, \ m_n \in &M \ for \ n=1,2,3,\ldots,n, \\
& but \ M \notin m_n.
\end{split}
\end{equation*}

Here $\in$ denotes a membership relation, where $M$ is at the level X, and $m_{n}$ are at the level Y.

By its axioms in one way or another, various modern set theory forbids that an item $M$ to be a member of an item $m$, given that one already has that $m$ as a member of $M$ (Jech 2006). The reason for such forbiddenness is due to above paradoxes such as Russell's paradox. The above relationship becomes trivial because such relationship arises directly from an axiom of modern set theories. Nonetheless, one cannot fail to notice that there exists a dissymetry between $m$ and $M$, whether the relationship is trivial in one theory and is not in another. In naive set theory, they are asymmetric in membership relationship, and in modern set theories, their relationship will mean that there is a hierarchy between $m$ and $M$: one is a $member$ of the other. Our conjecture simply states and emphasizes that there exists such dissymetry, and that the study of emergence is a study of such dissymetry. The relationship between the concept of emergence and set theoretic paradoxes and concepts may be an interesting topic of future studies.

\section{Brief history: `symmetry-breaking' as a physical example of novelty producing process}

In what follows, we present an example of a novelty producing process well studied in the physical sciences, that of `symmetry-breaking.' After all, symmetry-breaking is the principal example Anderson (1972) presented as the novelty producing process in his seminal paper:

\begin{quotation}
\noindent "In my own field of many-body physics, we are, perhaps, closer to our fundamental, intensive underpinnings than in any other science in which nontrivial complexities occur, and as a result we have begun to formulate a general theory of just how this shift from quantitative to qualitative differentiation takes place. This formulation, called the theory of `broken symmetry', may be of help in making more generally clear the breakdown of the constructionist converse of reductionism." (393)
\end{quotation}

We focus on \textit{spontaneously} broken symmetry (as opposed to \textit{explicitly} broken symmetry, where some symmetry-breaking term is explicitly added into the equations governing the system: for example, consider an external magnetic term toward a particular spatial direction). It is defined by Anderson (1984) in the following manner:

\begin{quotation}
\noindent "Although the equations describing the state of a natural system are symmetric, the state itself is not, because the symmetric state can become unstable towards the formulation of special relationships among the atoms, molecules, or electrons it consists of." (265)
\end{quotation}

Before we proceed to explain Anderson's definition, it is important to state the \textit{principle of minimum energy}: for a closed system with constant external parameters and entropy, the internal energy will decrease and the system will ultimately approach the state with a \textit{minimum energy} at equilibrium.  This is a restatement of the second law of thermodynamics, stating that in every real process the sum of the entropies of all participating bodies is increased. This principle is an empirical finding accepted as an axiom of physics. The fact that this is an axiom/law by itself without any derivation from other axiom/laws is important and will be emphasized later.

Now we will explain the terms in Anderson's definition. Let's assume that in our physical system of interest, that there exist governing equations which can describe the dynamics of the system and describe the possible states the system can possibly be in (e.g., in physics such governing equations for the system are often called Hamiltonian or Lagrangian of the system). For example, if such governing Hamiltonian/Lagrangian does not show any preference of the \textit{spatial orientation} for the system to be in (e.g. the equation does not tell the system to align toward top direction or bottom, or to point left or right, or toward you or against you), then the equations have symmetry with respect to the spatial orientation, and are said to be symmetric under the operation of \textit{translations}. The system, indeed, can take advantage of such symmetry and be in such a symmetric state. 

\begin{figure}[h!]
\centering
\includegraphics[width=0.8\textwidth]{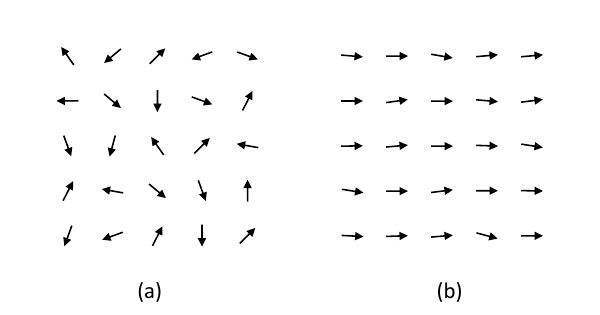}
\caption{Two systems of magnets expressed as an arrow are shown (the direction of the arrow meaning from S to N of the magnet). (a) The state of the system above Curie temperature, in which the magnets point to random directions. (b) The state of the system below Curie temperature, in which the magnets point to one direction.}
\label{fig1}
\end{figure}

In nature, we can observe the phenomena of ferromagnetism. For a ferromagnetic system (a system with very many ferromagnets) at high temperature, each individual component making up the system has random orientations like Figure 1 (a).  Indeed, for such ferromagnetic systems, the model equations we use to describe them have orientation symmetry.  The interaction terms in the equations do not have any preferences for the orientations.  However, at low temperatures, a novel phenomenon appears: \textit{the symmetry of the system suddenly breaks down}. Under a critical point called Curie temperature, there appears stable states with individual components all aligned along a particular direction, as shown in Figure 1 (b). Now the newly aligned states have the \textit{lower energy} compared to the symmetric random state; therefore, they become stable according to the principle of minimum energy, and the random state with symmetry becomes an unstable state, for it has higher energy.

What we observed here is one example of phase transition. Such phase transitions can happen not only in a system consisting of many magnets.  It can also happen in a system with many H2O molecules, He molecules, and super conducting materials, among many others. In a system of H2O molecules, under the critical point of zero degree Celsius, the system now turns into a crystal called ice, by aligning in a particular pattern of lattice-like configuration. Above that temperature, the system gains symmetry (and becomes water) and the molecules freely move without any alignment and configurations.

In summary, the symmetry of a large system can be very different from the symmetry of the Hamiltonian of the system. The Hamiltonian, describing the interactions and motions of the individual, does not tell what kind of symmetry the system will have. We needed another law to tell you that: the minimum energy principle. Thus, restating Anderson's definition of spontaneous broken symmetry, "although the equations describing the state of a natural system are symmetric, the state itself is not, because the symmetric state can become unstable towards the formulation of special relationships among the atoms, molecules, or electrons it consists of." (265) 

We summarize the traits of the spontaneous symmetry-breaking of a large system from the \textit{critical phenomena} point of view for the phase transition:

\begin{enumerate} [I.]
\item The system must be made of a very large number of individual components.
\item In our example, the components were homogeneous (this happens to be an unnecessary condition for phase transition: indeed, studies in physics tell us that much richer phenomena at the macro level are possible if the components are inhomogeneous).
\item There is a \textit{critical point} where the symmetry of the system begins to break.
\item Beyond the critical point, the Hamiltonian that was able to describe the macro-properties of the system loses its ability to describe.
\item We needed another law -- especially the minimum energy principle -- to describe the   
macro-properties/laws of broken symmetries in the large system.
\end{enumerate}

We emphasize that the law we required to explain the broken symmetry appears beyond the critical point for the large system, which is the minimum energy principle.  This is an axiom in physics, inspired by empirical findings.  By the definition of axiom, it does not require any other laws to derive itself.  In such manner, this exists independently from the microscopic laws. We can state that the macroscopic property/law of broken symmetry relies on the minimum energy principle that is not determined from the symmetries of the original Hamiltonian (arising out of the microscopic interactions). Therefore, the third essential claim of emergence, "novelty of emergent properties", that some properties/laws at the system level be novel and have the same status as the micro-level properties/laws, is satisfied. There may be cases that, upon close inspection, the microscopic interactions and laws that make up the Hamiltonian also rely on the minimum energy principle. Still, in such a case, both the macroscopic law and microscopic law will be at the level of status, and novel in the sense that the descriptions from the minimum energy principle at each level are different from each other.

\section{Thesis B: spontaneous symmetry-breaking as emergence example}

We summarize Anderson's argument for the spontaneous symmetry breaking being the example of emergence as the following conjecture:

\begin{quotation}
\noindent \textbf{Conjecture B (Anderson's Thesis)}: Spontaneous symmetry-breaking is a novel emergent property at the macro level, which cannot be explained from the laws/properties at the micro level.  
\end{quotation}

In other words, according to the Anderson's Thesis, detecting spontaneous symmetry-breaking is equivalent to detecting the emergence in the system.

As shown above, the broken-symmetry of a complex system is manifested as a phase transition of the system. Do all phase transitions accompany symmetry-breaking such that we can define a phase transition as the signature of a symmetry-breaking of a system? The answer is no. There are known examples of phase transitions which do not accompany symmetry-breaking. Still, insofar as a symmetry-breaking of a system involves a phase transition, studies of properties of which are common to phase transitions will give us information regarding symmetry-breaking as shown in the following example.

\section{Brief history: hysteresis and time symmetry-breaking}

We now focus on another phenomenon, \textit{hysteresis}. Again, it is a phenomenon associated with phase transition. In the broadest sense, hysteresis is a phenomenon where the current state of the system depends on its past history. The most popular example is again of the magnets (ferromagnetic materials), now in the presence of the external magnetic field.

For example, for a large ferromagnetic system made of many small magnets, consider the case where the temperature is low enough such that the aligning of the magnets reduces the energy, and such that in its natural circumstances the magnets are all aligned in one direction. Now we can apply an external magnetic field to the aforementioned system. As shown in Figure 2, when the external field is strong enough, all the magnets in the systems align in one direction, following the direction of the external field (A and B in Figure 2). When we reverse the direction of the magnetic field, the magnets in the system will now reverse its direction, and if the field is strong enough, all magnets will be aligned towards the opposite direction. During the process, the system itself acts as if it has inertia, and tends to stay where the system was at its previous stage. The arrows in the Figure 2 show such a history dependent path of the system. The dark area within the path is called hysteresis loop, and this area is the measure of the hysteresis for the system. Because of such path dependency, the system can be in either state C or D with exactly the same external magnetic field, temperature, pressure, etc.

\begin{figure}[h!]
\centering
\includegraphics[width=0.7\textwidth]{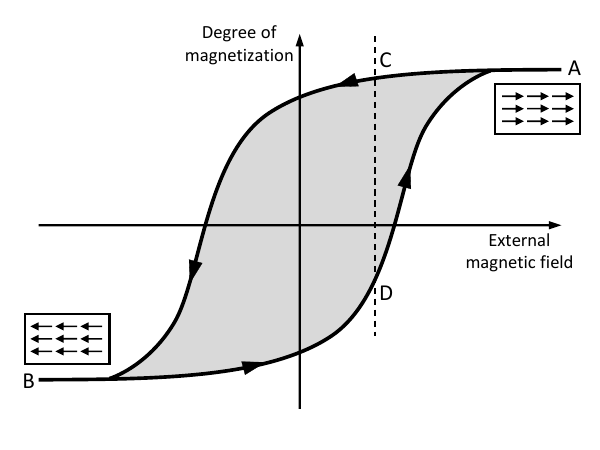}
\caption{A system of magnets under external magnetic field exhibiting a hysteresis loop is shown. At point A and B, the magnets are aligned in the direction of external magnetic field. When the external magnetic field is changed, the system will either go through state C or D depending on its previous state.}
\label{fig2}
\end{figure}

The Phenomenon of hysteresis is another example of symmetry breaking. Here, the broken symmetry is with respect to the dimension of time. Laws governing physical objects usually have symmetry with respect to the direction of time, as we find in Newton's law of motion describing the trajectories of moving objects. For example, when one throws a ball into the sky and makes a film of the trajectory, there is no way to know even if the film is projected in the time reversed order: Newton's law predicts that the motion of the ball going up is exactly opposite of the motion of the ball going down. So, if you project the film in the rewound fashion, it will be as if it is shown in the original time direction.

That is not the case with the phenomenon of hysteresis. When one begins at point A in the Figure 2, you must pass the upper path going through C to reach B. When one starts from B, one must pass the lower path through D to come back to A. If one films the action and projects it, it will be immediately noticeable if the film is shown in reversed time direction.

\section{Thesis C: hysteresis ans emergence example}

Yuri Mnyukh (2013) argues that the phenomenon of hysteresis is a universal feature (at least) in the phase transitions of condensed matters. Mnyukh argues that the mechanism of phase transition is \textit{nucleation and propagation} in most, if not all, cases, and it inherently features hysteresis as one of its characteristics. If Mnyukh's argument is correct, then hysteresis (thus, time symmetry-breaking) would be a common feature of condensed matter phase transition, including the spontaneous symmetry-breaking example given in Section 7. We summarize Mnyukh's argument as the following conjecture:

\begin{quotation}
\noindent \textbf{Conjecture C (Mnyukh's Thesis)}: Hysteresis (time symmetry-breaking) accompanies all phase transitions. 
\end{quotation}

\color{black}

\section{Curie’s Principle of symmetry properties, and relationship between emergence, membership broken-symmetry, and hysteresis}

We begin this section by restating three conjectures:

\begin{quotation}
\noindent \textbf{Conjecture A (New Emergentist Thesis)}: A system realizing hierarchical realism (having a broken-symmetry in membership relation) can have novel emergent properties at the macro level.
\medbreak
\noindent \textbf{Conjecture B (Anderson's Thesis)}: Spontaneous symmetry-breaking is a novel emergent property at the macro level, which cannot be explained from the laws/properties at the micro level.  \\
\medbreak
\noindent \textbf{Conjecture C (Mnyukh's Thesis)}: Hysteresis (time symmetry-breaking) accompanies all phase transitions, thus also the one caused by the spontaneous symmetry-breaking.
\end{quotation}

We point out that these conjectures are related to each other. Conjecture B gives an example and a way to confirm conjecture A: if there is a spontaneous symmetry-breaking, then there is an emergent phenomenon in the system. Note that the inverse might not be true. There can be other types of symmetry-breaking for an emergent system. Conjecture C is related to conjecture B. If a system shows a spontaneous symmetry-breaking, then it will exhibit a hysteresis and is an emergent system. Cojecture C can also be related to conjecture A. If hysteresis indeed accompanies all phase transitions, it will also accompany phase transitions caused by the symmetry-breaking of the system.

These propositions look at the problem of emergence from the perspective of symmetry breaking. The New Emergentist Thesis essentially becomes the question of whether symmetry-breaking in membership relation causes spontaneous symmetry-breaking (or time symmetry-breaking) in the system at the macro level, which cannot be explained from any other laws/properties at the micro level.
Upon answering such questions, we may take hints from principles stated by Pierre Curie, the very person who contributed to the study of systems in Section 8 and Section 9, and came up with the concept of Curie Temperature.
We recite Curie's principle of symmetry properties from Brading and Castellani (2005,  4):

\begin{quotation}
\noindent 1.1. When certain causes produce certain effects, the symmetry elements of the causes must be found in their effects.
\medbreak
\noindent 1.2. When certain effects show a certain dissymmetry, this dissymmetry must be found in the causes which gave rise to them.
\medbreak
\noindent 1.3. In practice, the converses of these two propositions are not true, i.e., the effects can be more symmetric than their causes.
\medbreak
\noindent 2. A phenomenon may exist in a medium having the same characteristic symmetry or the symmetry of a subgroup of its characteristic symmetry. In other words, certain elements of symmetry can coexist with certain phenomena, but they are not necessary. What is necessary, is that certain elements of symmetry do not exist. Dissymmetry is what creates the phenomenon.
\end{quotation}

Logically, 1.1 and 1.2 are the same, and 1.3 clarifies the statement. 2. reiterates and emphasizes the role of dissymmetric cause in the dissymmetric outcome. The effects may be more symmetric than the cause, because the dissymmetry does not necessarily have to be transferred from the cause to the effect. However, if there is a dissymmetry (broken-symmetry) in the effect, there must be a dissymmetry in the cause. Essentially, the claim of New Emergentist Thesis is that such cause of dissymmetry is not in the micro level laws/properties, but in the macro level laws/properties, which originating in turn from the dissymmetry in the membership relation between micro and macro.

Common sense and standard philosophy of physics probably suggest that when a dissymmetric state comes to exist from the symmetric initial state, we infer either (i) there was an initial asymmetry not registered in the symmetric initial state, or (ii) it is due to sheer indeterminism. Curie's principle dictates there must be some dissymetry in the cause. The New Emergentist Thesis argues that the dissymetry in the membership relation is enough of a cause, whether the system in question is deterministic or not.

These propositions look at the emergent phenomenon from the perspective of symmetries. Now then, can we prove them? Do we have enough evidences about these conjectures such that they can be considered an experimentally and/or theoretically proven thesis? We wish to discuss further this question in future works, also providing a clearer relationship between broken-symmetry and emergence in due course.

\section{Epilogue}

There is a fundamental question about the novelty of emergence related to the very nature of science. At its heart, science is crucially tied to an inductive mode of reasoning.  Every theory is based on natural observations, and the aim of every theory is to explain and predict the natural observations.  The validity of a theory is confirmed by experiments, and a theory is valid only if the next experiment correctly confirms it so. As Karl Popper puts it, every scientific theory should be falsifiable by the next experiment. There is Renormalization Group (RG) Theory in the field of statistical physics which may explain some examples of critical phenomena of phase transition. At the heart of the theory, there exists an iterative process called renormalization by which we draw out macroscopic variables from the microscopic ones. According to the theory, for the group of systems in the same category called universality class, the microscopic details of the systems does not matter in drawing out the macroscopic properties. It may very well be that the theory is self-consistent and without any loopholes, such that we can prove the novelty at the macroscopic level and therefore the existence of emergence with the theory. RG does not need to be a sole theory for such a task as explaining the mechanism for the emergence in question. We can construct many other theories, which explain natural phenomena fittingly; and at the same time explains the mechanism for the emergence under investigation. However, there is always a chance that the theory may be falsified by our next experiment. We need theories to observe and explain the phenomena of nature, and there is never a guarantee that \textit{any} of those theories will be a correct one tomorrow. In this respect, the proof of novelty and claim of emergence is always epistemic (with the possibility of being ontological).  But by practical means, in so far as our theory at hand is not wrong \textit{yet}, we can assume that the claim of emergence is ontological, at least given what we know today.

\pagebreak

\end{document}